\documentclass[aps,pre,reprint,10pt,superscriptaddress,showpacs]{revtex4-1}
\usepackage{amsfonts} 
\usepackage{amsmath}
\usepackage{amssymb}
\usepackage{graphicx}
\usepackage{subfigure}
\usepackage{color}
\newcommand*\xbar[1]{%
  \hbox{%
    \vbox{%
      \hrule height 0.5pt 
      \kern0.5ex
      \hbox{%
        \kern-0.1em
        \ensuremath{#1}%
        \kern-0.1em
      }%
    }%
  }%
} 
\begin{document}
\title{Stability and evolution of electromagnetic solitons in relativistic degenerate  laser plasmas}
\author{Sima Roy}
\email{simaroy031994@gmail.com}
\affiliation{Department of Mathematics, Siksha Bhavana, Visva-Bharati (A Central University), Santiniketan-731 235,  India}
\author{A. P. Misra}
\email{apmisra@visva-bharati.ac.in; apmisra@gmail.com}
\affiliation{Department of Mathematics, Siksha Bhavana, Visva-Bharati (A Central University), Santiniketan-731 235,  India}
 \begin{abstract}
 The dynamical behaviors of electromagnetic (EM) solitons formed due to  nonlinear interaction of linearly polarized intense laser light and relativistic degenerate plasmas are studied. In the slow motion approximation of relativistic dynamics, the evolution of weakly nonlinear  EM  envelope is described by the generalized nonlinear Schr{\"o}dinger (GNLS) equation with local and nonlocal nonlinearities. Using the Vakhitov-Kolokolov criteria, the stability of an  EM soliton solution  of the GNLS equation is studied. Different stable and unstable  regions  are demonstrated with the effects of soliton velocity, soliton 
 eigenfrequency, as well as the  degeneracy parameter $R=p_{Fe}/m_ec$, where $p_{Fe}$ is the   Fermi momentum and $m_e$ the electron mass, and $c$ is the speed of light in vacuum. It is found that the stability region shifts to an unstable one and is significantly reduced as one enters from the regimes of weakly  relativistic $(R\ll1)$ to ultrarelativistic   $(R\gg1)$ degeneracy of electrons. The analytically predicted results are in good agreement with the   simulation results of the GNLS equation. It is shown that the standing EM soliton solutions are  stable. However,     the moving solitons  can be stable or unstable depending on the values of soliton velocity, the eigenfrequency or the degeneracy parameter. The latter with  strong degeneracy $(R>1)$  can eventually lead to soliton collapse.  
\end{abstract}
\date{28 Oct 2020}
\maketitle
\section{Introduction}\label{sec-intro}
Extensive studies on the formation and dynamics of solitons  has been receiving renewed interests because of their fundamental importance in nonlinear sciences,  as well as  an important candidate for the emergence of    turbulence in nonlinear dispersive media (See, e.g., Ref. \onlinecite{misra2011} and references therein). 
Highly relativistic degenerate dense plasmas are believed to exist in compact astrophysical objects, e.g., in the interior of white dwarfs, neutron star  and magnetars  with particle number density ranging from $10^{26}$ cm$^{-3}$ to $10^{34}$ cm$^{-3}$. Such high-density degenerate plasmas may be directly created via ultraviolet or x-ray free electron lasers \cite{williams2019}.  Also, by varying the laser intensity, partially or fully degenerate plasmas can also be produced in the laboratory \cite{hayes2020}, which makes  laser produced plasmas to be useful recreating astrophysical plasmas in the laboratory. Thus, successful operations of lasers in laboratories  open up new possibilities to study the EM pulse penetration and to explore the subsequent nonlinear dynamics in degenerate dense plasmas under laboratory conditions.
\par
Recent investigations stipulate that  currently available laser intensity is as high as above $10^{24}$ W/cm$^{2}$ \cite{jeong2016}. In the laser-plasma interactions, the electrons (and also protons) at these high intensities may no longer be nonrelativistic but are forced to accelerate with velocities close to the speed of light in vacuum. Relativistic motion of these plasma particles strongly affects the dynamics of laser pulse propagation, especially when nonlinear effects, e.g., ponderomotive force come into the picture, and thereby leading to various interesting nonlinear phenomena  including the formation of relativistic EM solitons. The latter are localized structures associated with the density depletion of electrons in the ponderomotive force field of an intense laser beam of light. These solitons are usually formed behind the laser pulse and they propagate in the form of an envelope  carrying a large part of the laser pulse energy and thereby serving as a candidate for laser beam energy conversion.
\par 
The  nonlinear propagation EM solitons in different plasma environments has been investigated by a number of authors \cite{verheest2015,verheest2016,bersons2020,holkundkar2018}. Furthermore, the dynamics of EM solitons in relativistic plasmas in an idealized case of circular polarization has been extensively studied within the framework of one-dimensional relativistic fluid model and particle-in-cell (PIC) simulations \cite{lehmann2006,sundar2011,mikaberidze2015,berezhiani2015}. The theory has been revisited in the context of linearly polarized EM waves as well \cite{mancic2006,hadzievski2002,roy2020}. Recently, the   formation of standing EM solitons for circularly polarized EM waves in degenerate relativistic plasmas  has been studied by Mikaberidze \textit{et al.} \cite{mikaberidze2015}. They showed that EM solitons can be stable both in weakly relativistic and ultrarelativistic degenerate plasmas. However, the existence and the stability of moving EM solitons in the context of linearly polarized intense laser in relativistic degenerate plasmas has not been explored in details. 
\par 
In this paper, our aim is to study the dynamics of EM solitons that are formed due to the nonlinear interaction of linearly polarized intense laser beam of light and electron density perturbations   driven by the laser ponderomotive force in relativistic degenerate dense plasmas. We show that the existence and stability of EM solitons are significantly modified by the effects of electron degeneracy. The analytical results are also shown to be in good agreement with the mumerical simulation of the nonlinear evolution equation.    
\section{Dynamical equations}\label{sec-baeqs}
 The nonlinear interaction of linearly polarized finite amplitude intense laser pulse with longitudinal electron density perturbations that are driven by the laser ponderomotive force in a relativistic degenerate plasma can be described by the following set of  dimensionless equations  which are the EM wave equation, the electron continuity and  momentum equations and the Poisson equation in the Coulomb gauge  \cite{gratton1997,roy2020,misra2018}.  
 \begin{equation}
\left(\frac{\partial^2}{\partial t^2}- c^2\frac{\partial^2}{\partial z^2}\right) {A}_x= \frac{\omega_{pe}^2m_ec^2}{n_0H_e}n_e^2A_x, \label{eq-em-wave}  
\end{equation}
\begin{equation}
 \frac{\partial \left(\gamma_en_e\right)}{\partial t} +\frac{\partial}{\partial z} (\gamma_en_e {v}_{ez})=0,\label{eq-cont}
 \end{equation}
\begin{equation}
\frac{d}{dt}\left(\frac{\gamma_eH_ev_{ez}}{n_ec^2}\right)=e\left(\frac{\partial\phi}{\partial z}-\frac{1}{2}\frac{en_e}{\gamma_eH_e}\frac{\partial A_x^2}{\partial z}\right)-\frac{1}{\gamma_en_e}\frac{\partial P_e}{\partial z}, \label{eq-moment}
\end{equation}
 \begin{equation}  
 \frac{\partial^2 \phi}{\partial z^2}=4\pi e(\gamma_en_e-n_0), \label{eq-poisson}
\end{equation}
where $d/dt\equiv\partial/\partial t+ {\bf v}_j \cdot \nabla$, and $e$,  $n_e$ (with  its equilibrium value $n_0$ in laboratory frame),  $m_e$ and $v_{ez}$ are, respectively, the charge,  the number density,   the mass and $z$-component of the velocity of electrons. Since ions form the neutralizing background,   $\gamma_i=1$ and  so, $\gamma_i n_i=n_0$, the equilibrium number density of electrons and ions. Thus, the electron number density in the Laboratory frame with its equilibrium and perturbation parts may be written as $n_L\equiv\gamma_en_e=n_0+n'_L$. 
 Furthermore, $c$ is the speed of light in vacuum, $\omega_{pe}$ is the electron plasma frequency, $A_x$ is the $x$-component of the vector potential, $\phi$ is the electrostatic potential, $P_e$ is the electron degeneracy pressure at zero temperature, given by, \cite{chandrasekhar1935}  
\begin{equation}
 \begin{split}
 &\left(P_e,{\cal E}_e\right)=\frac{m_e^4c^5}{3\pi^2\hbar^3}\left[f(R),~R^3\left(1+R^2\right)^{1/2}-f(R)\right],\\
 &f(R)=\frac18\left[ R\left(2R^2-3\right)\left(1+R^2\right)^{1/2}+3\sinh^{-1}R\right],
 \end{split} \label{pressure}
\end{equation}  
where $\hbar=h/2\pi$ is the reduced Planck's constant, $R=p_{Fe}/m_ec=\hbar\left(3\pi^2n_e\right)^{1/3}/m_ec$ is the dimensionless degeneracy parameter, and  
 $H_e\equiv P_e+{\cal E}_e=n_em_ec^2\sqrt{1+R^2}$ is the enthalpy per unit volume measured in the rest frame of each element of the fluid with ${\cal E}_e$ denoting the total energy density, i.e., ${\cal E}_e=m_en_ec^2+\bar{\epsilon}_e$ and $\bar{\epsilon}_e$    the internal energy of the fluid.
Also, $\gamma_e$ is the relativistic factor, given by, \cite{misra2018}
\begin{equation}
 \gamma_e=\sqrt{\frac{1+ e^2n_e^2  A^2_x/H^2_e}{1-v_{ez}^2/c^2}}. \label{eq-gamma}
\end{equation} 
For the slow-motion approximation of relativistic dynamics of electrons, i.e., 
$en_eA_x/H_e \sim o(\epsilon)$;    $\phi,~n_{e1},~v_{ez}\sim o(\epsilon^2)$, where $0<\epsilon<1$ is a scaling parameter,  Eqs. \eqref{eq-em-wave} to \eqref{eq-poisson} can be reduced to the following coupled equations \cite{misra2018}. 
 \begin{equation}\label{eq-basic-A}
\left(\frac{\partial^2}{\partial t^2}-\frac{\partial^2}{\partial z^2}+1\right)A+(1-\delta_{e})(N-\alpha A^2)A=0,
\end{equation}  
\begin{equation}\label{eq-basic-N}
\left(\frac{\partial^2}{\partial t^2}-\delta_{e}\frac{\partial^2}{\partial z^2}+1\right)N=\frac{1}{2}(1-\delta_{e})\frac{\partial^2 A^2}{\partial z^2},
\end{equation} 
where $N\equiv N_{e1}/n_0=\gamma_en_e/n_0-1$ is the dimensionless longitudinal electron density perturbation and $A\equiv \eta_e eA_x/m_ec^2$ is the dimensionless  vector potential along the $x$-axis. The  space and time coordinates are  normalized according to $t\rightarrow t\sqrt{\eta_e}\omega_p $, $z\rightarrow z\sqrt{\eta_e}\omega_p/c$.  Also, $\alpha=\eta_e^2/2$, $\delta_e=(1-\eta_e^2)/3$ and $\eta_e=1/\sqrt{1+R_0^2}$  with $R_0=\hbar\left(3\pi^2n_0\right)^{1/3}/mc$ denoting a measure of the strength of plasma degeneracy, i.e., $R_0\ll1$ corresponds to  the weakly relativistic degenerate plasma, whereas  $R_0\gg1$ is referred to as ultra-relativistic degenerate  plasmas.
\par For the evolution and stability  of EM solitons in relativistic degenerate plasmas, it appears much more difficult to solve the fully relativistic one-dimensional fluid model   due to the generation of higher harmonics of linearly polarized incident laser pulse. However,  a more convenient approach  is   to study the   coupled equations \eqref{eq-basic-A} and \eqref{eq-basic-N} instead. 
So, we introduce a slowly varying complex wave envelope in the form. Here, we note that   in contrast to  circular polarization, the linearly polarized EM waves have odd harmonics for the vector potential $A$ and even harmonics for the electron density perturbation $N$. Thus, we have
 \begin{equation}\label{eq-expan-A-N}
\begin{split}
&A=\frac{1}{2}\left(ae^{-it}+a^{*}e^{it}\right),\\
&N=N_{0}+\frac{1}{2}\left(N_{2}e^{-i2t}+N_{2}^{*}e^{i2t}\right), 
\end{split}
\end{equation}
where the asterisk denotes the complex conjugate of the corresponding physical quantity. 
Substituting the expansions in Eq. \eqref{eq-expan-A-N} for  $A$ and $N$ into Eq. \eqref{eq-basic-N}, and collecting the zeroth and second harmonic terms $(\sim e^{-i2t})$, we obtain the following envelopes for $N_{0}$ and $N_{2}$.
\begin{equation}\label{eq-N0-N2}
\begin{split}
&N_{0}=\frac{1}{4}(1-\delta_{e})(|a|^{2})_{zz},\\
&N_{2}=-\frac{1}{12}(1-\delta_{e})(a^{2})_{zz}.
\end{split}
\end{equation}
Next, substituting Eq. \eqref{eq-expan-A-N} into  Eq. \eqref{eq-basic-A}, and  collecting the first harmonic terms $(\sim e^{-it})$, we obtain the following   equation  for the EM wave amplitude $a$ [For details see Appendix \ref{sec-app}. 
\begin{equation}\label{eq-nls}
\begin{split}
i\frac{\partial a}{\partial t}+\frac{1}{2}(a)_{zz}&+\frac{3}{8}\alpha(1-\delta_{e})|a|^{2}a-\frac{1}{8}(1-\delta_{e})^{2}(|a|^{2})_{zz}a\\
&+\frac{1}{48}(1-\delta_{e})^{2}(a^{2})_{zz}a^{*}=0.
\end{split}
\end{equation}
Equation  \eqref{eq-nls} has the form of a generalized nonlinear Schr{\"o}dinger (GNLS) equation with local (cubic) as well as nonlocal  (derivative) nonlinearities.  It is   noticed that  both the cubic and nonlocal nonlinear coefficients are significantly modified by the  effects of  the relativistic electron degeneracy pressure.  It is to be noted that in the limit of $R_0\rightarrow0$, Eqs. \eqref{eq-N0-N2} and \eqref{eq-nls} assume   the same form as Eqs. (7) and (8) in Ref. \cite{hadzievski2002} after replacing   $A$ by $a$. However, there are some disagreements with the factor $1/2$ in the expression for $N_2$ [Eq. \eqref{eq-N0-N2}] and in the last term of Eq. \eqref{eq-nls}. This may be due to some   sign mismatch in the coefficient of  $({A^2}/{2})A$   [Eq. \eqref{eq-basic-A}] with the similar term in Ref. \cite{hadzievski2002}. The   cubic nonlinearity   should be correctly as    $({A^2}/{2})A$ or $({a^2}/{2})a$    instead of $-({a^2}/{2})a$ as in Ref. \cite{hadzievski2002}. Similar nonlinear term with   the   correct positive sign can be found  in some other previous investigations, e.g., Eq. (29) of  Ref. \cite{gratton1997}.      
\par 
We look for a localized stationary solution of Eq. \eqref{eq-nls} in the form of a moving soliton $a=\rho(\xi)\exp[i\theta(\xi)+i\lambda^{2}t]$ where  $\xi=z-v_0t$  and $v_0$ is the soliton velocity in the moving frame of reference. Thus, substituting this solution into Eq. \eqref{eq-nls}, we obtain the following equations for the soliton phase and the amplitude.
\begin{equation}\label{eq-soltn-phse}
\begin{split}
&\theta_{\xi\xi}\rho\left[1+\frac{1}{12}(1-\delta_{e})^{2}\rho^2\right]\\
&+\theta_{\xi}\rho_{\xi}\left[2+\frac{1}{3}(1-\delta_{e})^{2}\rho^{2}\right]-2v_0\rho_{\xi}=0,
\end{split}
\end{equation}
\begin{equation}\label{eq-soltn-amp}
\begin{split}
\rho_{\xi\xi}&-\frac{5}{12}(1-\delta_{e})^{2}\frac{\rho}{\zeta}\rho_{\xi}^{2}=\frac{\rho}{\zeta}\left(2\lambda^{2}-2v_0\theta_{\xi}+\theta_{\xi}^2\right)\\
&+\frac{\rho^3}{\zeta}\left[\frac{1}{6}(1-\delta_{e})^{2}\theta_{\xi}^{2}-\frac{3}{4}\alpha(1-\delta_{e})\right] =0,
\end{split}
 \end{equation}
  where $\zeta(\rho)=1-(5/12)(1-\delta_{e})^{2}\rho^{2}$. 
Using the same boundary conditions, namely $\rho,~\rho_{\xi},~\rho_{\xi\xi}\rightarrow0$ as $\xi\rightarrow\pm\infty$,  the integration of Eq. \eqref{eq-soltn-phse} approximately yields $\theta(\xi)_{\xi}=v_0$, while that of  Eq. \eqref{eq-soltn-amp} gives
\begin{equation}\label{eq-rho_xi}
\rho_{\xi}^{2}=\frac{\rho^2}{\zeta}\left(2\lambda^2-v_0^2-\left[\frac{3}{8}\alpha(1-\delta_{e})-\frac{1}{12}(1-\delta_{e})^{2}v_0^{2}\right]\rho^2\right). 
\end{equation}
Further integration of Eq. \eqref{eq-rho_xi} yields a soliton solution in the following
  implicit form.
\begin{equation}\label{eq-soliton}
\begin{split}
\pm \xi =& \sqrt{\frac{10(1-\delta_e)}{\Delta}}\ln\frac{\sqrt{\zeta}+\sqrt{\zeta-\zeta_0}}{\sqrt{\left\vert\zeta_0\right\vert}}\\
&-\frac{1}{2\sqrt{2\lambda^2-v_0^2}}\ln\frac{\sqrt{1-\rho^2/\rho_0^2}+\sqrt{\zeta}}{\left\vert\sqrt{1-\rho^2/\rho_0^2}-\sqrt{\zeta}\right\vert}, 
\end{split}
\end{equation}
where  $\zeta_0(\rho_0)=1-(5/12)(1-\delta_{e})^{2}\rho_0^{2}$, $\Delta=9\alpha-2(1-\delta_{e})v_0^{2}$ and 
\begin{equation}\label{eq-rho0}
\rho_{0}^{2}=\frac{24(2\lambda^{2}-v_0^{2})}{\Delta(1-\delta_{e})} 
\end{equation}
is the squared  amplitude (maximum) of the linearly polarized moving EM soliton with the reduced eigenfrequency $\Lambda\equiv\lambda^2-v_0^2$. In particular, for $v_0=0$, one can obtain the standing soliton solution of the GNLS equation \eqref{eq-nls}.  
 \begin{figure*}
\includegraphics[scale=0.36]{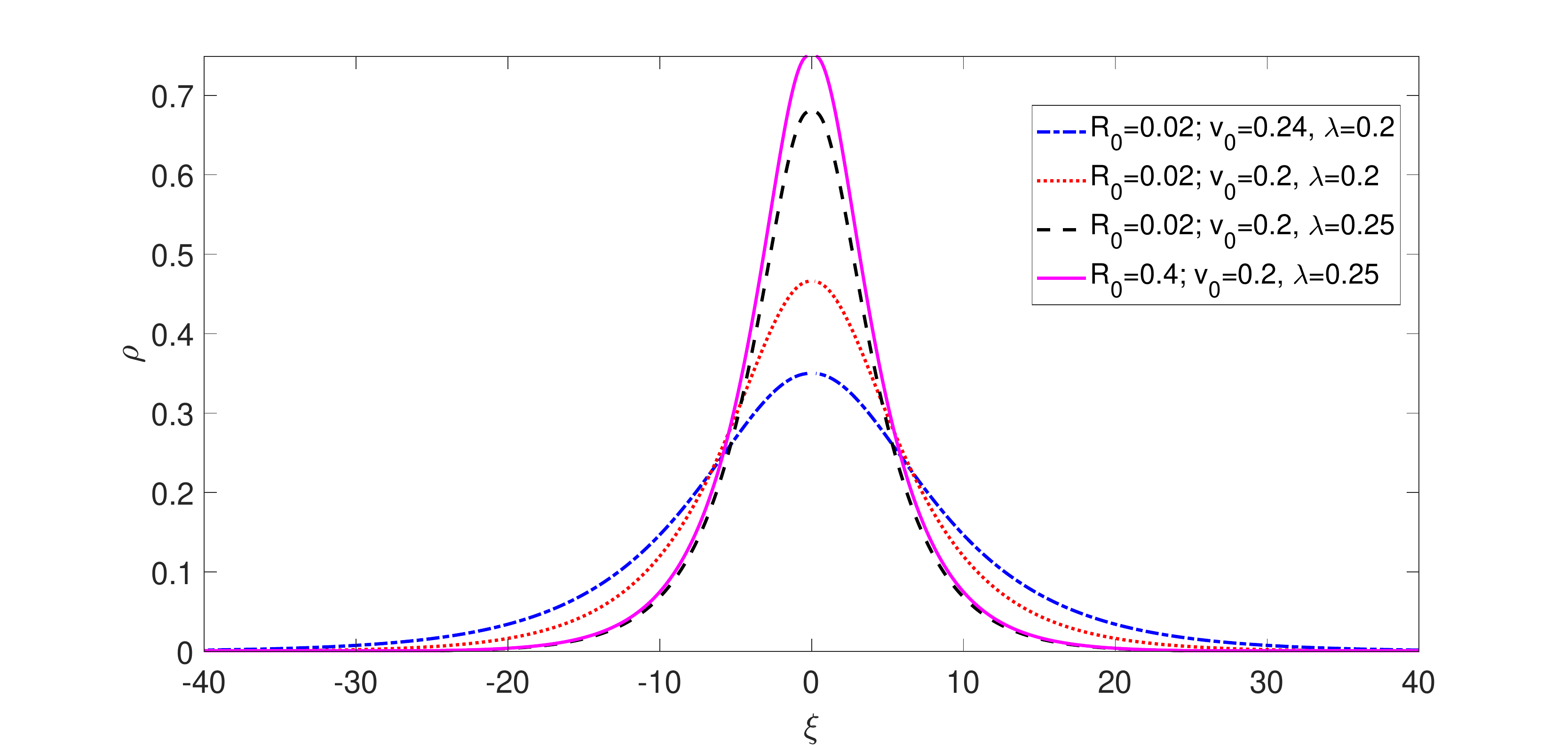}
\caption{A profile of the EM soliton, given by Eq \eqref{eq-soliton}, is shown  for different values of the relativistic degeneracy parameter $R_{0}$, the soliton eigenfrequency $\lambda$, and the soliton velocity $v_0$ as in the legend. }
\label{fig-soliton}
\end{figure*}
In general, Eq. \eqref{eq-soliton} describes a two-parameter family of solutions of Eq. \eqref{eq-nls} which are symmetric about the origin, and while combined together generates the  form a soliton. As an illustration, we plot the soliton solution [Eq. \eqref{eq-soliton}] as shown in Fig. \ref{fig-soliton} for different values of the parameters $v_0$, $\lambda$ and $R_0$. It is seen that  for a fixed value of $R_0$ and $\lambda$ but with an increasing value of the soliton velocity,  the amplitude decreases while the soliton profile (width) broadens. An opposite feature is observed with increasing value of $\lambda$, i.e., the soliton amplitude increases   but the width decreases. However, when the relativistic degeneracy effect is more pronounced, an enhancement  of both the amplitude and width of the soliton is seen to occur. In order to justify the physics behind it we consider the   the small amplitude limit of solitons in which the nonlocal (ponderomotive) nonlinear terms may be  neglected so that one recovers the  cubic NLS equation from Eq. \eqref{eq-nls} as
\begin{equation}
i\frac{\partial a}{\partial t}+P\frac {\partial^2 a} {\partial z^2}+Q|a|^2a=0, \label{eq-nls-redu}
\end{equation}
where $P=1/2$ and $Q=(3/8)\alpha(1-\delta_e)$. Equation \eqref{eq-nls-redu} has a traveling wave solution of the form $a\sim \sqrt{a_0}~\text{sech} [(z-v_0t)/w]$, where $wa_0=\sqrt{2|P/Q|}$. Physically, the relativistic degeneracy pressure of electrons provides the wave dispersion (which increases with increasing values of $R_0$) quite distinctive from the dispersion due to separation of charged particles. So, depending on the dispersion and nonlinear effects, the soliton amplitude and width can increase or decrease. Since $P$ appears as a constant due to its normalization, the wave amplitude/width can increase or decrease according to when the value of the nonlinear coefficient $Q$ decreases or increases.    It is noted that the values of   $Q$ decrease  with increasing values of $R_0$. Thus, it may be likely that the soliton amplitude and width can increase with increasing values of $R_0$.    However, when the wave amplitude is not small, the ponderomotive nonlinear effects can intervene the dynamics and compete with the cubic nonlinearity which may result into an increase of the wave amplitude and/or width or some other nonlinear phenomena including collapse.   
\section{Stability Analysis} \label{sec-stbi-anlys}
In order to examine  the stability of the moving EM soliton, we follow the Vakhitov-Kolokolov stability criteria \cite{vakhitov1973}. According to the criteria, solitons are stable under   the longitudinal perturbations  if 
\begin{equation} \label{eq-stb-cond-P0}
\frac{dP_{0}}{d\lambda^{2}}>0, 
\end{equation}
  where $P_{0}$ is the soliton photon number defined by   
 \begin{equation}\label{eq-P}
P=\int |a|^{2} dz.
\end{equation}
 An expression for $P_{0}(\lambda)$ can be obtained for  the soliton solution  \eqref{eq-soliton}  as
\begin{equation}\label{eq-photon}
 P_{0}(\lambda)=\frac{\sqrt{6}}{\sqrt{\Delta(1-\delta_{e}) }} 
\left( \rho_{0}+\sqrt{\frac{3}{5}}\frac{\zeta_0}{1-\delta_{e}} \ln\frac{1+ \sqrt{1-\zeta_0}}{\left\vert 1- \sqrt{1-\zeta_0}\right\vert}\right)
\end{equation}
\begin{figure*}
\includegraphics[scale=0.36]{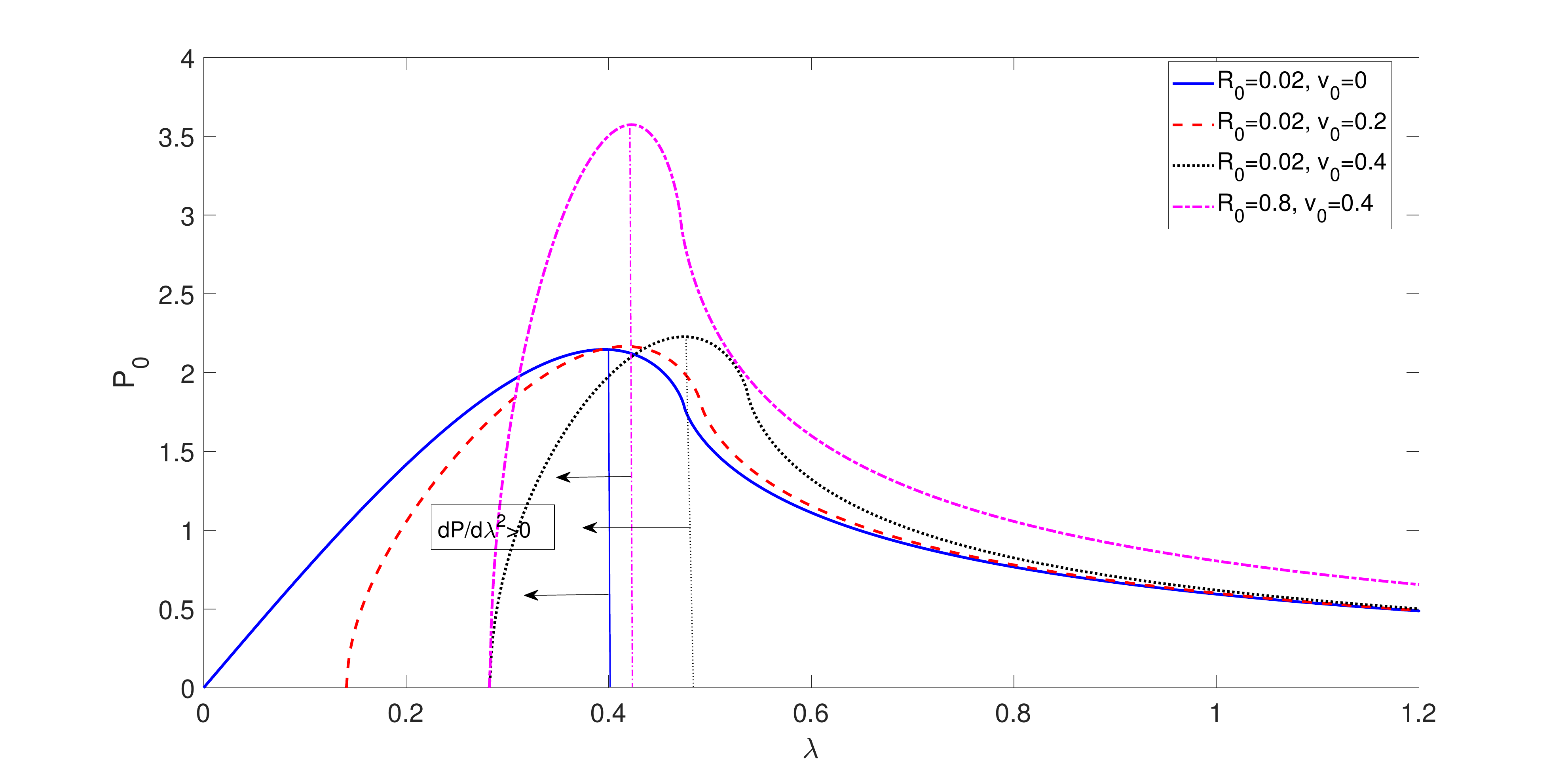}
\caption{Photon number $P_{0}(\lambda)$, given by Eq. \eqref{eq-photon}, is plotted against the eigenfrequency $\lambda$ to show the stable $(\lambda<\lambda_s; dP_0/d\lambda^2>0)$ and unstable $(\lambda>\lambda_s; dP_0/d\lambda^2<0)$ regions for different values of the relativistic degeneracy parameter $R_{0}$ and the soliton velocity $v_0$ as in the legend. The straight lines are drawn to indicate the corresponding threshold values  $\lambda_s$ of $\lambda$ at which $P_0$ reaches a  maximum value.    }
\label{fig-photon}
\end{figure*}
\par 
So, according to the condition  \eqref{eq-stb-cond-P0}, the moving EM soliton \eqref{eq-soliton} turns out to be stable in the region $\lambda<\lambda_{s}$, where $\lambda_s$ is some  instability threshold value of $\lambda$ at which $P_0$ reaches a local maximum and above which $dP_0/d\lambda^2<0$. An   analytic expression of $\lambda_{s}$ cannot be determined in its explicit form. However, we try to find its values numerically for different values of the degeneracy parameter $R_0$ and the soliton velocity $v_0$.
\par 
 The profiles of the curves of $P_0(\lambda)$, as shown in Fig. \ref{fig-photon},    predicts the existence of stable $(\lambda<\lambda_s)$ and unstable regions  $(\lambda>\lambda_s)$. It is found that an increase of the soliton velocity $v_0$ shifts the instability threshold $\lambda_s$ towards larger values of it than that for the standing soliton with $v_0=0$ (see the solid and dashed curves). However, $\lambda_s$ can shift  towards its lower values if  the degeneracy parameter $R_0$ is increased (see the dotted and dash-dotted curves). It follows that  the stability domain $\lambda<\lambda_{s}$ of EM solitons may be significantly reduced in the regime  of  strongly or ultrarelativistic degenerate plasmas $(R_0\gg1)$. We mention that  the Vakhitov-Kolokolov criterion predicts only a linear stability of EM solitons involving the exponentially growing or decaying modes.  So, it may not   predict  about the subsequent nonlinear evolution of unstable EM  envelopes or about the stability of  localized structures with arbitrary profiles. In general, the  GNLS equation  \eqref{eq-nls}  can admit, apart from the stationary soliton solution,  the soliton collapse \cite{litvak1978} and  long-lived relaxation oscillations around the stable soliton amplitude  due to cubic nonlinearity, and perhaps some other dynamical states due to the presence of nonlocal nonlinearities \cite{hadzievski2002,pelinovsky1996}. An alternative  stability criterion  for the solitons     can also be formulated in terms of the Hamiltonian and photon number interrelation (See, e.g., Ref. \onlinecite{mancic2006}). We, however,   skip this analysis here, instead look for some other regimes for the existence  and stability of solitons by revisiting the Vakhitov-Kolokolov stability criterion and the constraints on the  parameters $\lambda$ and $v_0$ in the analytical soliton
solution  \eqref{eq-soliton}.  The limits of $\lambda$ and $v_0$ are given by
\begin{equation}\label{eq-cond-stb1}
2\lambda^2-v_0^2>0, ~~ \zeta\equiv1-5\rho^{2}\left(1-\delta_{e}\right)^{2}/12>0. 
\end{equation}       
 On more limitation on the parameter $v_0$ is given by 
 \begin{equation} \label{eq-cond-stb2}
 \Delta\equiv 9\alpha-2(1-\delta_e)v_0^2>0, ~\rm{i.e.,}~ v_0<v_s\equiv \sqrt{9\alpha/2(1-\delta_e)}.
 \end{equation}
 An estimation reveals that for $0<(\alpha,~\delta_e,~v_0)<1$, the condition $\Delta>0$ holds for $0\leq R_0\lesssim1.36$ and $\Delta<0$ for $R>1.36$. 
 \par
\begin{figure*}
\includegraphics[scale=0.36]{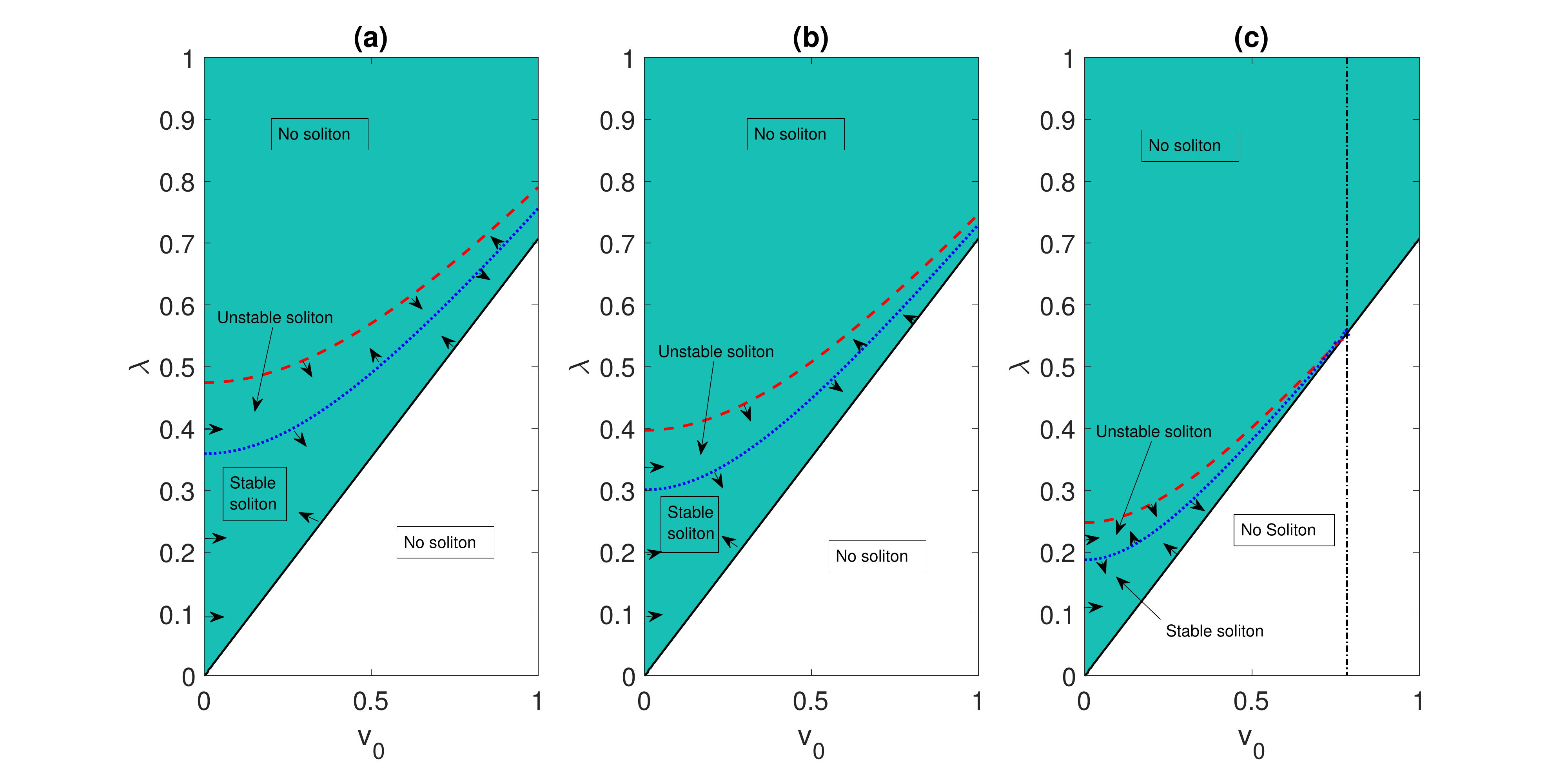}
\caption  { The regions for the existence of EM solitons and their stability are shown in the $(v_0,\lambda)$-plane for different values of the relativistic degeneracy parameter $R_{0}$:   (a)  $R_{0}=0.02$,   (b)  $R_{0}=0.8$ and (c) $R_0=2$.  The solid, dashed and dotted lines, respectively,  represent the contour plots of $2\lambda^{2}-v^{2}=0$, $\zeta\equiv 1-(5/12)\rho^{2}(1-\delta_{e})^{2}=0$ and $ dP_0/d\lambda^2=0$.  In the subplot (c), the  dash-dotted  line represents a threshold value of $v_0$: $v_s=0.78$ above which no soliton solution exists as $\Delta\equiv 9\alpha-2(1-\delta_e)v_0^2<0$ there. Note that the value of $v_s$ decreases with increasing values of $R_0>1.36$. No such threshold exists in the region $0\leq v_0\leq1$, $0\leq\lambda\leq1$ when $R_0$ lies in $0\leq R_0\lesssim1.36$. }
\label{fig-stableregion}
\end{figure*}

 In what follows, considering all the stability conditions, namely Eqs. \eqref{eq-stb-cond-P0}, \eqref{eq-cond-stb1} and \eqref{eq-cond-stb2} imposed on the parameters $\lambda$ and $v_0$, 
we   define the regions of the soliton existence and stability in $(v_0,\lambda)$-plane  as shown in Fig. \ref{fig-stableregion}.  We find that above the dashed curve of $\zeta\equiv 1-5\rho^{2}(1-\delta_{e})^{2})/12=0$, no analytic soliton solution exists as the conditions are not satisfied there. However, one can obtain numerically a   soliton solution of Eq. \eqref{eq-nls} there. Such a discrepancy in this particular region  occurs due to an initial small  phase $[\theta(\xi)]$  difference   between these   approximate analytical and numerical soliton  solutions. Furthermore, below the solid curve of $2\lambda^{2}-v^{2}=0$ no localized solution (analytical or numerical) can exist.  The existence region of solitons is separated by a dotted curve.  Between the dashed and the dotted curves is the region where $dP_{0}/d\lambda^{2}<0$, but both $2\lambda^{2}-v^{2}$ and $\zeta>0$, and so in this region  only unstable soliton solution exists. However,    the stable soliton region is in between the solid and dotted curves where all the conditions for soliton stability  are satisfied. From the subplots (a) to (c)  it is noticed that as we gradually enter from the regimes of weak, moderate to strong relativistic degenerate plasmas (by increasing the value of the degeneracy parameter $R_{0}$) both the stable and unstable regions get significantly compressed. Also, a stable region shifts to an unstable one as $R_0$ increases. Furthermore,  subplot (c) shows that a threshold value $v_s$ of $v_0$ appears  when   $R_0>1.36$. In fact, for $R_0\gg1$, no stable or unstable region may be found in the $(v_0,\lambda)$-plane. This may be the limitation of the linear stability analysis which may not   correctly predict the existence   and the stability regions of moving solitons, especially   in  the regime of ultrarelativistic degeneracy.
\section{Simulation results}\label{sec-simul}
\begin{figure*}
\includegraphics[scale=0.36]{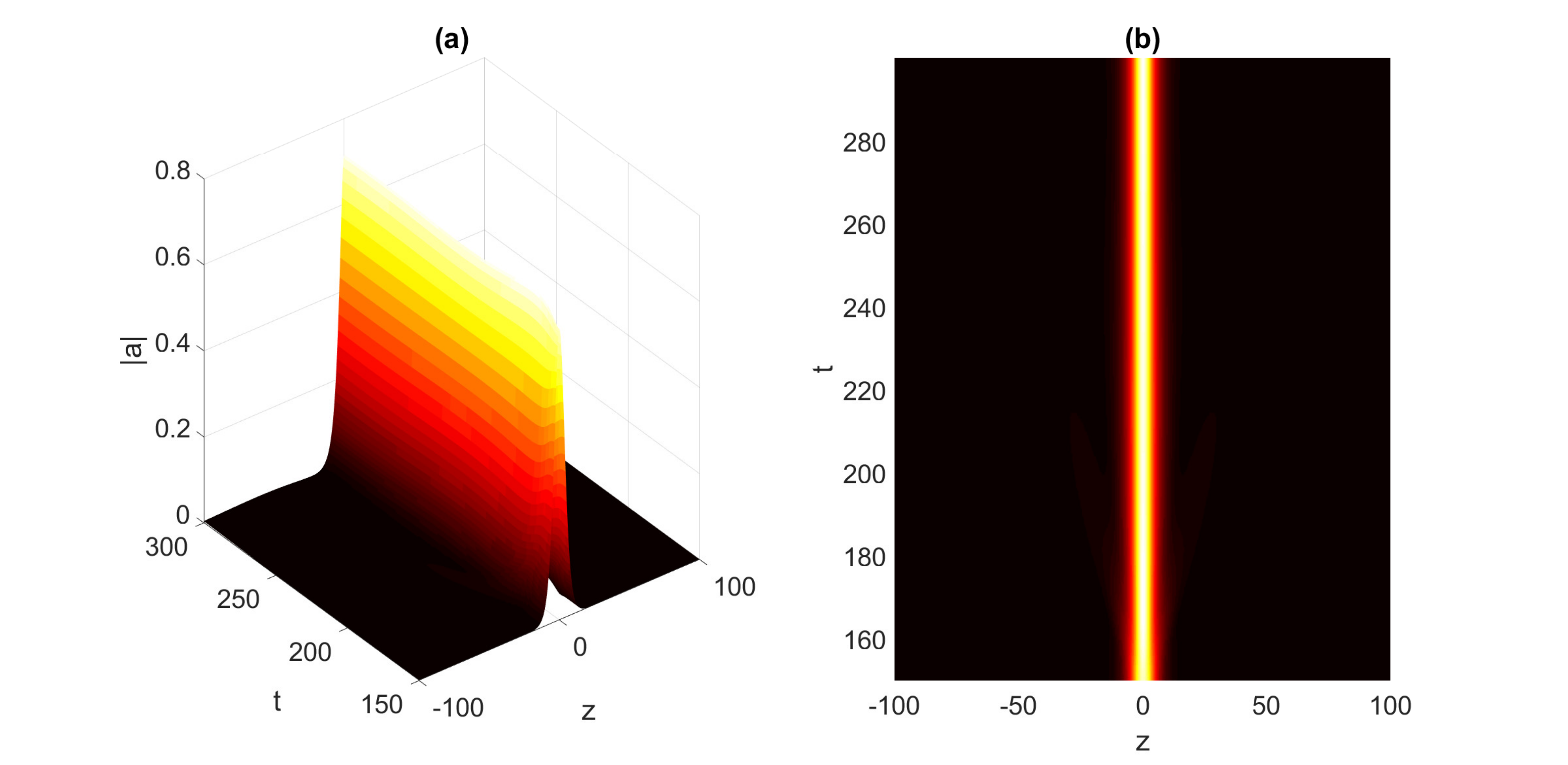}
\includegraphics[scale=0.36]{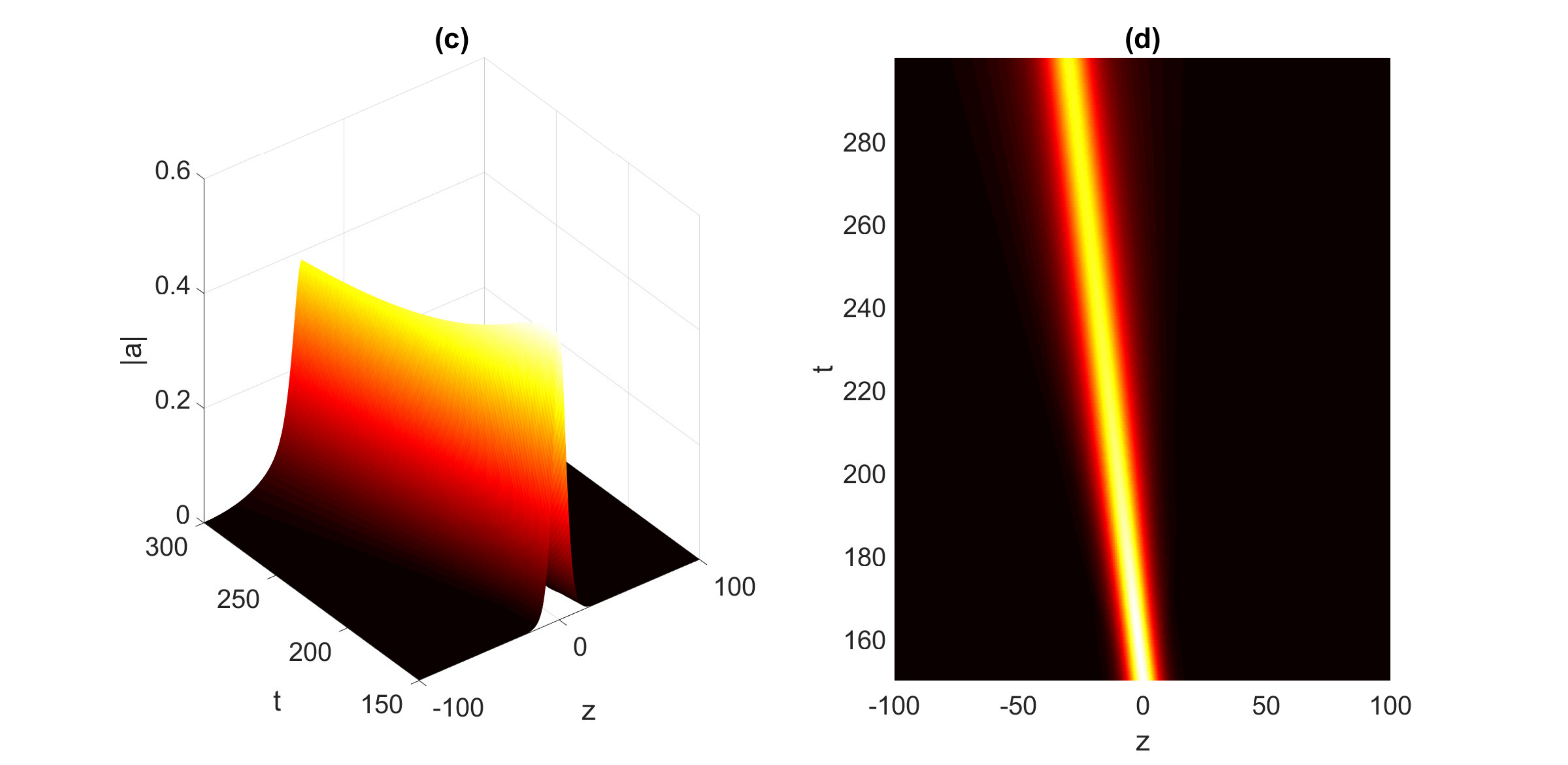}
\includegraphics[scale=0.36]{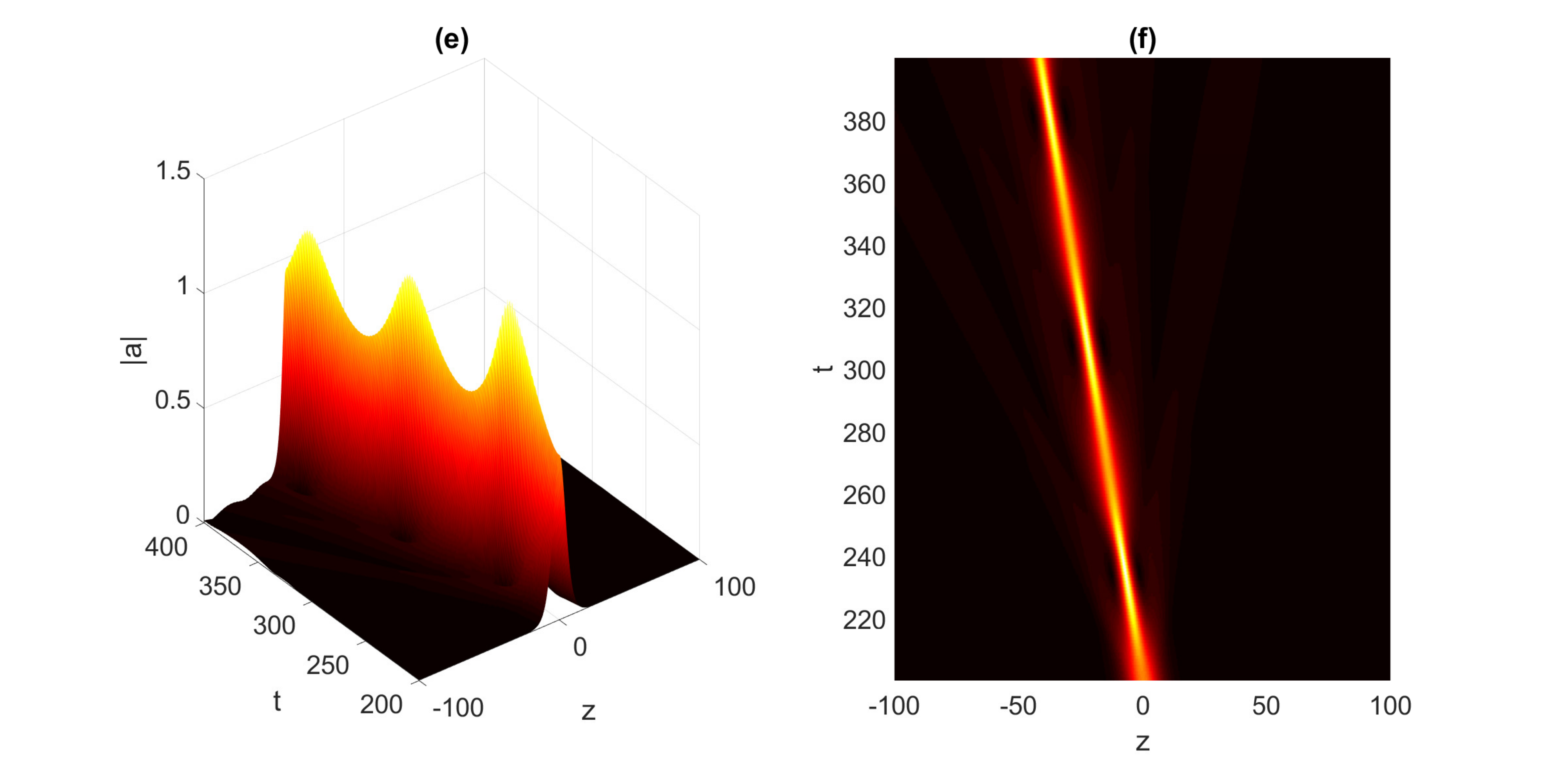}
\caption  {Spatio-temportal evolution of the generalized nonlinear Schrödinger (GNLS) equation \eqref{eq-nls} in different dynamical regimes for a fixed value of the degeneracy parameter $R_0=0.02$. Left panels are the surface plots of soliton solutions, while the right panels represent the corresponding contour plots. Subplots  (a)   and  (b)  show an evolution of  standing soliton  in the stable region  with soliton velocity $v_0=0$,   the soliton eigenfrequency $\lambda=0.207$, the amplitude $a_0=0.68$ and the soliton photon number $P_0=1.46$; Subplots (c) and (d)  a moving soliton in the stable region with an increasing velocity $v_0=0.2$, the same  $\lambda=0.207$  but different $a_0=0.67$ and $P_0=1.12$; Subplots      (e) and (f) a moving soliton in the stable region with the same $v_0=0.2$ as in [(c) \& (d)] but different $a_0=0.72$, $\lambda=0.26$ and $P_0=1.55$.}
\label{fig-evolution1}
\end{figure*}

\begin{figure*}
\includegraphics[scale=0.36]{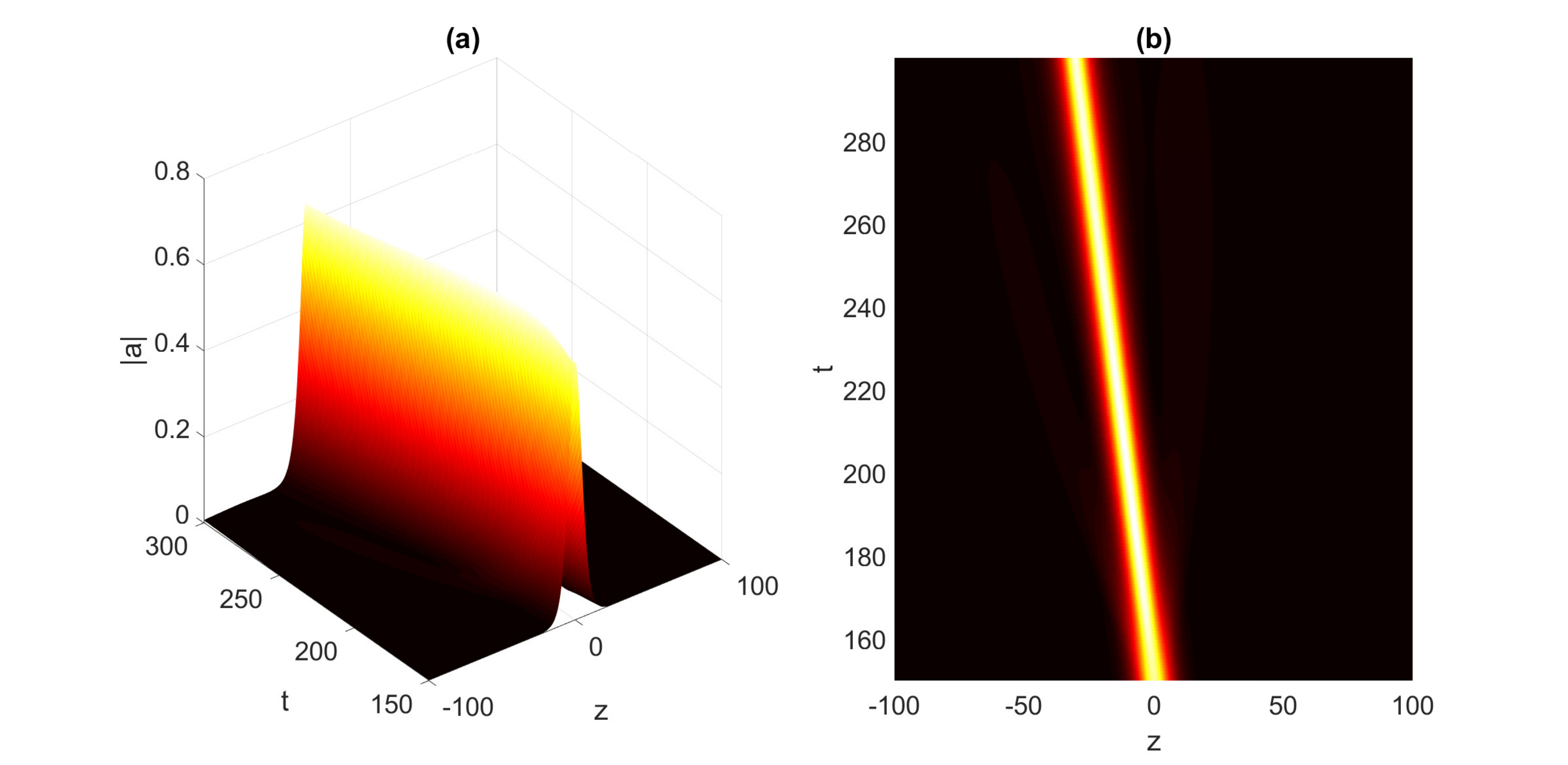}
\includegraphics[scale=0.36]{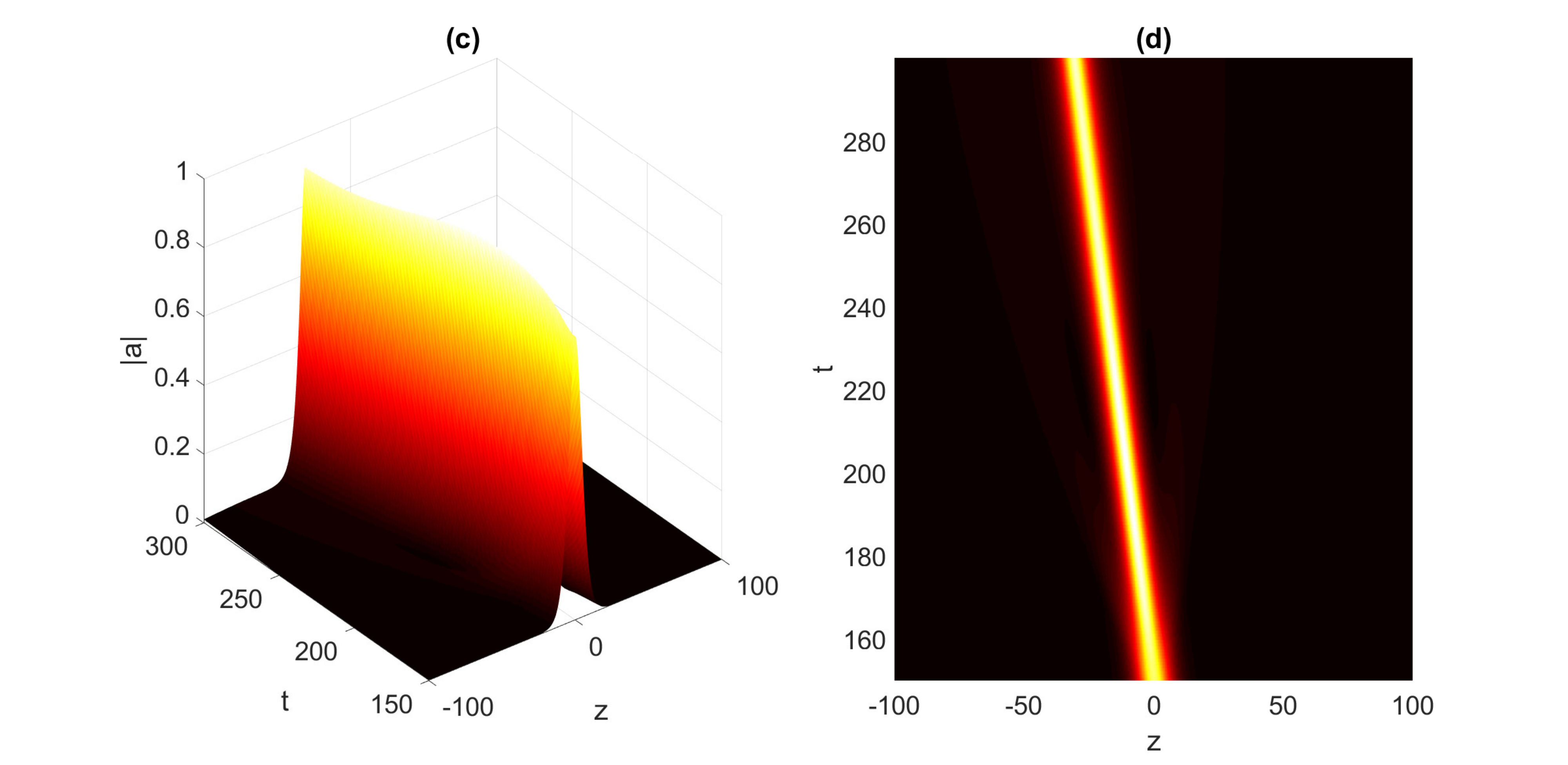}
\includegraphics[scale=0.36]{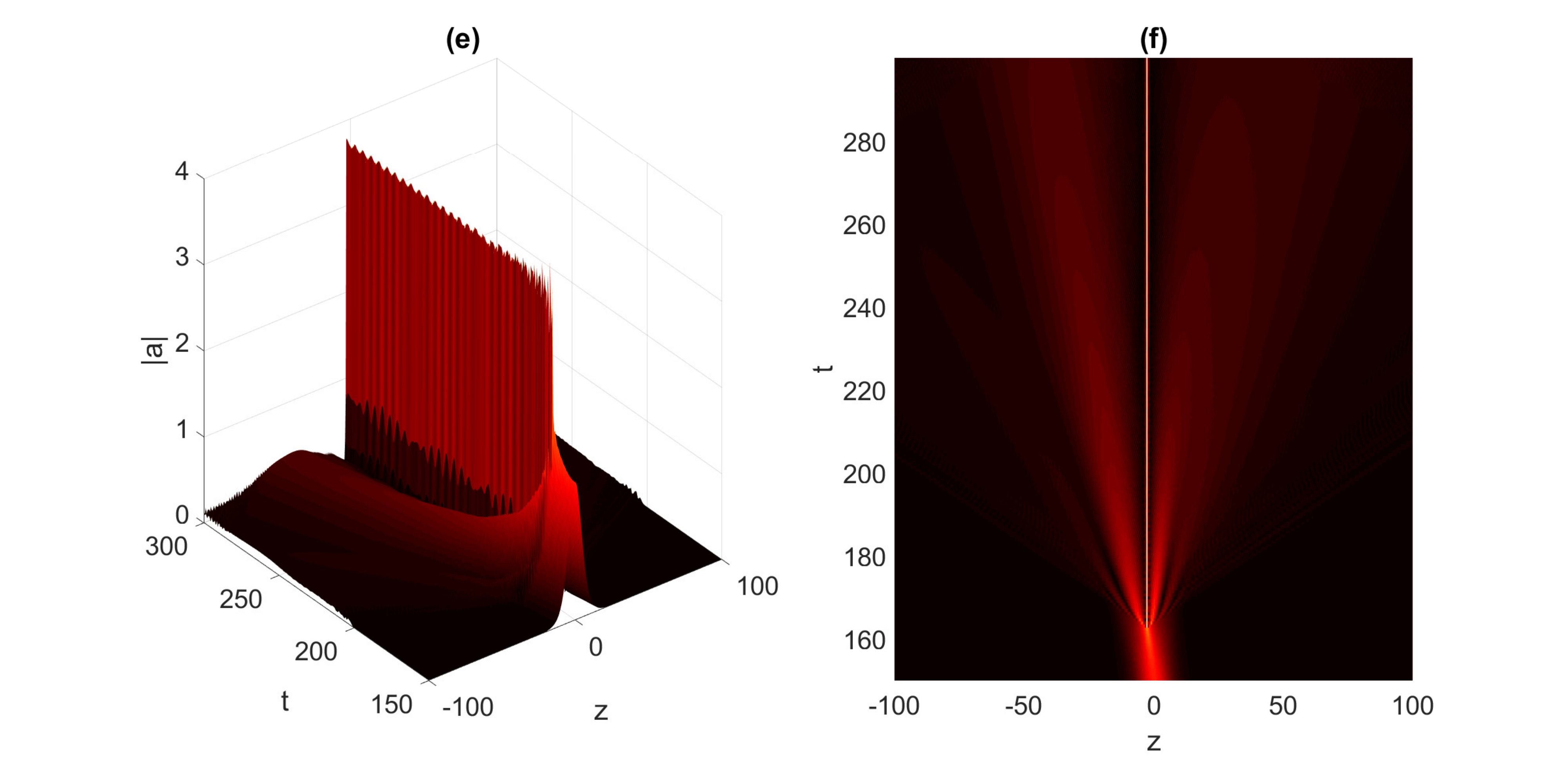}
\caption  {Spatio-temportal evolution of the generalized nonlinear Schr{\"o}dinger (GNLS) equation \eqref{eq-nls} for different values of the degeneracy parameter $R_0$:  $R_0=0.02$ in the stable region [subplots (a) and (b)], $R_0=0.8$ in the stable region [subplots (c) and (d)], and $R_0=2$ in the unstable region [subplots (e) and (f)]. The other parameter values are fixed at $v_0=0.2$ and $\lambda=0.23$. }
\label{fig-evolution2}
\end{figure*}

So far we have obtained the  possible regimes for the existence and stability of EM solitons based on an analytic stationary soliton solution of the GNLS equation \eqref{eq-nls}.   Next, we examine these   regimes by a direct numerical simulation of Eq. \eqref{eq-nls}. To this end,  we use  the Runge-Kutta scheme with a time step $dt=0.001$ and with an initial condition in the form of a soliton: $a(\xi,0)\sim a_0~\text{sech}^2(\xi/5)\exp(-i v_0\xi)$. The soliton evolution after time $t=300$ is shown in Figs. \ref{fig-evolution1} and \ref{fig-evolution2}  for different values of $v_0$, $\lambda$ and $R_0$  relevant for stable and unstable regions (\textit{cf}. Fig. \ref{fig-stableregion}) as predicted  in Sec. \ref{sec-stbi-anlys}.   
Here, we note that the initial condition may not be a solution to the GNLS equation \eqref{eq-nls}. However, as time goes on, the initial pulse radiates and   the nonlinear and dispersion effects intervene to evolve it as stable or unstable solitons. 
 Numerical simulation reveals that initially launched soliton \eqref{eq-soliton} with parameters $v_0$, $\lambda$ and $R_0$ in the stable region remains stable for a long time.  However, the stable or unstable behaviors may be changed if one considers the parameter values slightly above or below the stable or unstable region.  
\par 
 Figure \ref{fig-evolution1} shows the evolution of EM solitons in the weakly relativistic regime $(R_0=0.02)$. It is found that  in the stable region,   the   standing soliton $(v_0=0)$ with amplitude $a_0=0.68$,   $\lambda=0.207$ and photon number $P=1.46$ oscillates around the stable equilibrium  with a frequency close to the plasma oscillation frequency and so the soliton propagation remains stable for a long time [subplots (a) and (b)].  However,   the moving soliton with the same $\lambda=0.207$ but with an increased velocity $v_0=0.2$ and  reduced amplitude $a_0=0.67$  in the stable region travels towards the downstream and exhibits decay of its amplitude [subplots (c) and (d)]. Physically, for a fixed $\lambda$, as the soliton velocity increases, its frequency of  oscillation decreases. This results into a diminution of the photon number from $P=1.46$ to $P=1.12$, and thereby leading to a decay  of   the soliton amplitude.    However, as time goes on it  relaxes towards a corresponding stable soliton. On the other hand, retaining the soliton speed at $v_0=0.2$,  but   increasing its amplitude to  $a_0=0.72$ and the frequency to $\lambda=0.26$, we find that both the soliton eigenfrequency and the photon number   $(P_0=1.55)$  increase. As a result,   though the soliton   evolves   with  long-lived oscillating behaviors of the breather type, its  amplitude grows and it may be prone to instability [subplots (e) and (f)].  
A slight deviation from the stable equilibrium  with an increase of $\lambda$ or the initial perturbation with an increased amplitude $a_0$ for a fixed soliton velocity  can lead  to an  aperiodic  growth of the amplitude  and thereby  the onset of collapse.   
 \par 
 In order to examine the effects of the relativistic degeneracy  on the soliton dynamics, we consider  three different values of $R_0$, namely $R_0=0.02$,   $R_0=0.8$ and  $R_0=2$ to define, respectively, the weak, moderate and strong degeneracy of electrons. 
 With reference to Fig. \ref{fig-stableregion} (which predicts that the soliton stable region can shift  to an unstable one by an increasing value of $R_0$)   we   see from Fig. \ref{fig-evolution2} that for some fixed values of $\lambda=0.23$ and $v_0=0.2$, as $R_0$ increases  the soliton amplitude grows, it loses its stability and eventually collapses. Subplots (a) and (b) show that in the weakly relativistic regime of electron degeneracy, a moving soliton with an amplitude $a_0=0.6$ in the stable region travels towards downstream with preserving its  profile, i.e., the soliton remains stable for a longer time.  The solitons also remain  stable even in the limit of $R_0\rightarrow0$, i.e., when there is no electron degeneracy.   If the initial soliton profile is considered   with an increased amplitude with $a_0=0.82$   in the regime of moderate degeneracy with $R_0=0.8$, the soliton amplitude grows but it evolves about the stable equilibrium until it remains in the stable region [subplots (c) and (d)]. However, as the parameter $R_0$ is further increased to $R_0=2$, the moving soliton with amplitude $a_0=1.6$ falls in the unstable region. In this situation, solitons can not travel undistorted with a constant velocity. Its amplitude aperiodically grows and eventually collapses  [subplots (e) and (f)].  
 \par
 We note that when the electron degeneracy effect is small,  i.e.,  $R_0\ll1$,   one finds $\eta_e\sim1$ and $\delta_e\ll1$ for which the coefficients of the GNLS equation \eqref{eq-nls} appear as constants. In this case, one recovers the similar results as in  Ref. \cite{hadzievski2002}   with no degeneracy of electrons.  However, for  moderate or large values  of $R_0$, as the parameter increases,   the magnitudes of both the cubic and nonlocal nonlnearities tend to decrease, which results into an enhancement of the soliton amplitude and/or width.    This is expected, since in absence of the nonlocal terms, the GNLS equation \eqref{eq-nls} admits a soliton  solution with amplitude/width   proportional to $\sqrt{|P/Q|}$, where $P$ corresponds to the group velocity dispersion and $Q$ the cubic nonlinear effects. However, in presence of the nonlocal nonlinearities, not only the solitons grow with higher values of $R_0$, a rapid aperiodic growth in  amplitude creates a collapsing soliton  until   the amplitude reaches a critical value,  and we can no longer observe the  dynamic behaviors.   
 The subplot \ref{fig-evolution2}(e)   shows that when the values of $v_0$, and $\lambda$ fall slightly outside the stable region [\textit{cf} Fig. \ref{fig-stableregion}(c)], the soliton amplitude grows to become unstable, and eventually collapses. From the subplot  \ref{fig-evolution2}(f), it may be estimated that the soliton collapse   starts to occur after time $t\sim160$, and within the time interval $0<t<160$, the soliton amplitude remains of the order of unity.       
 Thus,   our analytical predictions for the existence and stability of EM solitons agree with the numerical results  that an increase of the soliton velocity shifts the instability threshold  towards larger values,  and  the strong degeneracy effects can shift  the steady sate  dynamics of solitons   to an unstable one which  results into wave collapse. 

 \section{Conclusion} 
We have studied the stability and dynamical evolution of electromagnetic (EM) solitons that are formed due to nonlinear interactions of linearly polarized intense laser light and relativistic degenerate dense plasmas in the framework of the generalized nonlinear Schr{\"o}dinger (GNLS) equation with local and nonlocal nonlinearities. The latter appear due to the laser driven ponderomotive force, and the generation of  odd and even harmonics for the vector potential and the electron density perturbation respectively.  An analytical moving soliton solution of the GNLS equation  and the corresponding soliton photon number $P$ are derived each in a closed form. 
\par 
A linear stability analysis of the soliton solution is performed according to the   Vakhitov-Kolokolov stability criteria. Different stable and unstable  regions  are demonstrated in the plane of the soliton velocity $v_0$ and the eigenfrequency $\lambda$ for different values of the degeneracy parameter $R_0$. It is found that the stability region shifts to an unstable one and is significantly reduced as $R_0$ increases from    the regimes of weak to strong relativistic degeneracy.  However, both the stable and unstable regions get significantly shrunk when $R_0\gg1$,i.e., in the limit of ultrarelativistic degeneracy, and eventually no feasible region can be found. This may be  the limitation of the  Vakhitov-Kolokolov stability criteria which may not predict the stability region for   $R_0\gg1$.  The stability analysis shows that the moving EM solitons in the weakly relativistic regime are stable with the stability region shifting towards smaller amplitudes in comparison with the standing soliton. However, as one enters from the weak to strong degenerate regime, the perturbation grows, i.e., the soliton stability region shifts towards larger amplitudes, and eventually the soliton collapses at higher value of $R_0>1$.  Furthermore, it is found that for an isolated soliton with a constant photon number, an increase of the soliton velocity results into a reduction of the maximum amplitude and broadening of the soliton profile.  Numerical simulation results of the GNLS equation are found to be in good agreement with our analytical predictions for the existence and  stability of EM solitons.
\par  It is to be noted that the determination of different dynamical regimes of EM solitons in the parameter space  $(P,\lambda)$ discussed above is important for understanding the low-frequency process of the formation of stable relativistic solitons behind the laser pulse  inside the photon condensate \cite{hadzievski2002}.  However, the detailed analysis of the regions in the parameter space requires    additional analytical and numerical study which is beyond the scope of the present work.
\par 
To conclude, the results should be useful for understanding the interactions of linearly polarized highly intense laser pulses with relativistic degenerate dense plasmas and their experimental verification as such experiments are going on with new generation of intense lasers, as well as  the characteristics of x-ray pulses emanating from compact astrophysical objects.   
 \section*{Acknowledgments}
This work was  supported by  the Science and Engineering Research Board (SERB), Govt. of India with  Sanction  order no. CRG/2018/004475   dated 26 March 2019.
\section*{ DATA AVAILABLITY STATEMENTS} Data sharing is not applicable to this article as no new data were created or analyzed in this study.
\appendix  
\section{} \label{sec-app}
 
Here, we give some details of the derivation of the GNLS equation \eqref{eq-nls}.  
\par 
 We consider the perturbation  expansions  for  $A$ and $N$  as
\begin{equation}\label{eq-exp-AN}
\begin{split}
&A=\frac{1}{2}\left(ae^{-it}+a^{*}e^{it}\right),\\
&N=N_{0}+\frac{1}{2}\left(N_{2}e^{-i2t}+N_{2}^{*}e^{i2t}\right). 
\end{split}
\end{equation}
Substituting Eq. \eqref{eq-exp-AN} into Eq. \eqref{eq-basic-N} we get
\begin{equation}
\begin{split}
&\left(\frac{\partial^2}{\partial t^2}-\delta_{e}\frac{\partial^2}{\partial z^2}+1\right)\left(N_{0}+\frac{1}{2}N_{2}e^{-i2t}+\frac{1}{2}N_{2}^{*}e^{i2t}\right)\\
&=\frac{1}{2}(1-\delta_{e})\frac{\partial^2}{\partial z^2}\left[\frac{1}{2}\left(ae^{-it}+a^{*}e^{it}\right)\right]^{2},
\end{split}
\end{equation}
or, 
\begin{equation} \label{eq-App1}
\begin{split}
&-2N_{2}e^{-i2t}-2N_{2}^{*}e^{i2t}+N_{0}+\frac{1}{2}N_{2}e^{-i2t}+\frac{1}{2}N_{2}^{*}e^{i2t}\\
&=\frac{1}{8}(1-\delta_{e})\left[(a^{2})_{zz}e^{-i2t}+2 (|a|^{2})_{zz}+(a^{*2})_{zz}e^{i2t} \right] 
\end{split}
\end{equation}
Collecting the zeroth and second harmonic terms $(\sim e^{-i2t})$, which appear due to self-interactions of waves, from both sides of Eq. \eqref{eq-App1}, we obtain the following expressions for $N_{0}$ and $N_{2}$.
\begin{equation} \label{eq-N0N2}
\begin{split}
N_{0}=\frac{1}{4}(1-\delta_{e})(|a|^{2})_{zz},\\ 
N_{2}=-\frac{1}{12}(1-\delta_{e})(a^{2})_{zz}.  
\end{split}
\end{equation}
Next, substituting Eq. \eqref{eq-exp-AN} into the wave equation \eqref{eq-basic-A} and using  Eq. \eqref{eq-N0N2} we obtain
\begin{equation}
\begin{split}
&\left(\frac{\partial^2}{\partial t^2}-\frac{\partial^2}{\partial z^2}+1\right)\left[\frac{1}{2}\left(ae^{-it}+a^{*}e^{it}\right)\right]\\
&+(1-\delta_{e})\left[N_{0}+\frac{1}{2}\left(N_{2}e^{-i2t}+N_{2}^{*}e^{i2t}\right)\right.\\
&\left.-\alpha \left(\frac{1}{2}\left(ae^{-it}+a^{*}e^{it}\right)\right)^2\right]\frac{1}{2}\left(ae^{-it}+a^{*}e^{it}\right)=0,
\end{split}
\end{equation}
or,
\begin{equation} \label{eq-App2}
\begin{split}
&\frac{1}{2}\frac{\partial^2a}{\partial t^2}e^{-it}-i\frac{\partial a}{\partial t}e^{-it}-\frac{1}{2}ae^{-it}+\frac{1}{2}\frac{\partial^2a^{*}}{\partial t^2}e^{it}+i\frac{\partial a^{*}}{\partial t}e^{it}\\
&-\frac{1}{2}a^{*}e^{it}-\frac{1}{2}(a)_{zz}e^{-it}-\frac{1}{2}(a^{*})_{zz}e^{it}\\
&+\frac{1}{2}ae^{-it}+\frac{1}{2}a^{*}e^{it}+\left[\frac{1}{4}(1-\delta_{e})^{2}(|a|^{2})_{zz}-\frac{1}{24}(1-\delta_{e})^{2}(a^{2})_{zz}e^{-i2t}\right.\\
&\left.-\frac{1}{24}(1-\delta_{e})^{2}(a^{*2})_{zz}e^{i2t}\right.\\
&\left.-\frac{1}{4}\alpha(1-\delta_{e})\left(a^{2}e^{-i2t}+2|a|^{2}+a^{*2}e^{i2t}\right)\right]\frac{1}{2}\left(ae^{-it}+a^{*}e^{it}\right)=0
\end{split}
\end{equation}
Collecting the first harmonic terms $(\sim e^{-it})$ from both sides of Eq. \eqref{eq-App2}  we get 
\begin{equation}\label{eq-App3}
\begin{split}
&\frac{1}{2}\frac{\partial^2a}{\partial t^2}-i\frac{\partial a}{\partial t}-\frac{1}{2}a-\frac{1}{2}(a)_{zz}+\frac{1}{2}a\\
&+\frac{1}{8}(1-\delta_{e})^{2}(|a|^{2})_{zz}a-\frac{1}{48}(1-\delta_{e})^{2}(a^{2})_{zz}a^{*}\\
&-\frac{1}{8}\alpha(1-\delta_{e})a^{2}a^{*}-\frac{1}{4}\alpha(1-\delta_{e})|a|^{2}a=0.
\end{split}
\end{equation}
For slowly varying wave envelopes, the terms involving $  \partial^2/ \partial t^2~(\ll1)$  may be neglected. Finally, we obtain from Eq. \eqref{eq-App3} the following wave equation for $a$.
\begin{equation}
\begin{split}
i\frac{\partial a}{\partial t}+\frac{1}{2}(a)_{zz}&+\frac{3}{8}\alpha(1-\delta_{e})|a|^{2}a-\frac{1}{8}(1-\delta_{e})^{2}(|a|^{2})_{zz}a\\
&+\frac{1}{48}(1-\delta_{e})^{2}(a^{2})_{zz}a^{*}=0.
\end{split}
\end{equation}


\begin{thebibliography}{50}  
\bibitem{misra2011} A. P. Misra, and S. Banerjee, Phys. Rev. E {\bf 83}, 037401 (2011). 
\bibitem{williams2019} G. O. Williams, H.-K. Chung, S. K{\"u}nzel, V. Hilbert, U. Zastrau, H. Scott,  S. Daboussi,  B. Iwan, A. I. Gonzalez, W. Boutu \textit{et al.}, Phys. Rev. Research {\bf1}, 033216 (2019).
\bibitem{hayes2020} A. C. Hayes, M. E. Gooden, E. Henry, G. Jungman, J. B. Wilhelmy, R. S. Rundberg, C. Yeamans, G. Kyrala, C. Cerjan, D. L. Danielson  \textit{et al.}, Nat. Phys. {\bf 16}, 432  (2020).
\bibitem{jeong2016} T. M. Jeong and J. Lee ``Generation of High-Intensity Laser Pulses and their Applications" DOI: 10.5772/64526.
\bibitem{verheest2015} F. Verheest, Phys. Scr. {\bf90}, 068002 (2015).
\bibitem{verheest2016} F. Verheest, Phys. Scr. {\bf 91}, 025603 (2016).
\bibitem{bersons2020} I. Bersons, R. Veilande, and O. Balcers, Phys. Scr. {\bf95}, 025203 (2020).
\bibitem{holkundkar2018} A. R. Holkundkar and G. Brodin, Phys. Rev. E {\bf 97}, 043204 (2018).
 \bibitem{lehmann2006} G. Lehmann, E. W. Laedke, and K. H. Spatschek {\bf13}, 092302 (2006). 
\bibitem{sundar2011} S. Sundar, A. Das, V. Saxena, P. Kaw, and A. Sen, Phys. Plasmas {\bf18}, 112112 (2011).
\bibitem{mikaberidze2015} G. Mikaberidzea and  V. I. Berezhiania, Phys. Lett. A {\bf 379}, 2730  (2015). 
\bibitem{berezhiani2015} V. I. Berezhiani, N. L. Shatashvili, and N. L. Tsintsadze, Phys. Scr {\bf90}, 068005 (2015).
\bibitem{mancic2006} A. Mancic, L. Hadzievski, and M. M. Skoric, Phys. Plasmas {\bf 13}, 052309 (2006).
\bibitem{hadzievski2002} Lj. Hadzievski, M. S. Jovanovic, M. M. Skoric, and K. Mima, Phys. Plasmas {\bf 9}, 2569 (2002).
\bibitem{roy2020} S. Roy, D. Chatterjee, and A. P. Misra, Phys. Scr. {\bf 95}, 015603 (2020).
  \bibitem{gratton1997} F. T. Gratton, G. Gnavi, R. M. O. Galvao, and L. Gomberoff, Phys. Rev. E
{\bf 55}, 3381 (1997).
\bibitem{misra2018} A. P. Misra and D. Chatterjee, Phys. Plasmas {\bf 25}, 062116 (2018).
 \bibitem{chandrasekhar1935} S. Chandrasekhar, Mon. Not. R. Astron. Soc. {\bf 95}, 207 (1935).
 \bibitem{vakhitov1973} N. G. Vakhitov and A. A. Kolokolov, Izv. Vyssh. Uchebn. Zaved. Radiofizika {\bf 16}, 1020 (1973).
\bibitem{litvak1978} A. G. Litvak and A. M. Sergeev, JETP Lett. {\bf27}, 517 (1978).
\bibitem{pelinovsky1996} E. Pelinovsky, V. V. Afanasijev, and Y. S. Kivshar, Phys. Rev. E 53,
1940 (1996).

 




\end{thebibliography}
\end{document}